# Keywords Extraction and Sentiment Analysis using Automatic Speech Recognition

Rachit Shukla

Electrical Engineering (2017-2021)

Punjab Engineering College, Chandigarh - 160012

## Abstract

Automatic Speech Recognition (ASR) is the interdisciplinary subfield of computational linguistics that develops methodologies and technologies that enables the recognition and translation of spoken language into text by computers. It incorporates knowledge and research in linguistics, computer science, and electrical engineering fields. Sentiment analysis is contextual mining of text which identifies and extracts subjective information in the source material and helping a business to understand the social sentiment of their brand, product or service while monitoring online conversations. According to the speech structure, three models are used in speech recognition to do the match: Acoustic Model, Phonetic Dictionary and Language Model. Any speech recognition program is evaluated using two factors: Accuracy (percentage error in converting spoken words to digital data) and Speed (the extent to which the program can keep up with a human speaker). For the purpose of converting speech to text (STT), we will be studying the following open source toolkits: CMU Sphinx and Kaldi. The toolkits use Mel-Frequency Cepstral Coefficients (MFCC) and I-vector for feature extraction. CMU Sphinx has been used with pre-trained Hidden Markov Models (HMM) and Gaussian Mixture Models (GMM), while Kaldi is used with pre-trained Neural Networks (NNET) as acoustic models. The n-gram language models contain the phonemes or pdf-ids for generating the most probable hypothesis (transcription) in the form of a lattice. The speech dataset is stored in the form of .raw or .wav file and is transcribed in .txt file. The system then tries to identify opinions within the text, and extract the following attributes: Polarity (if the speaker expresses a positive or negative opinion) and Keywords (the thing that is being talked about). For text processing, we will be studying the following unsupervised model algorithms and APIs: TextBlob, NLTK, TextRank, TopicRank, YAKE, and TF-IDF.

## Abbreviations

| | |
|---|---|
| ASR | Automatic Speech Recognition |
| STT | Speech to Text |
| MFCC | Mel Frequency Cepstral Coefficient |
| GMM | Gaussian Mixture Model |
| HMM | Hidden Markov Model |
| NLTK | Natural Language Toolkit |
| YAKE | Yet Another Keyword Extractor |
| TF - IDF | Term Frequency - Inverse Document Frequency |
| NLP | Natural Language Processing |
| API | Application Programming Interface |
| FFT | Fast Fourier Transform |
| ANN | Artificial Neural Network |
| CER | Character Error Rate |
| WER | Word Error Rate |
| SVM | Support Vector Machine |
| DNN | Deep Neural Network |
| WFST | Weighted Finite State Transducer |

# 1 INTRODUCTION

## 1.1 Background

Over the last few years, almost all the top notch organisations and companies have started to work around with speech recognition along with natural language processing (NLP). There are commercial systems such as Google Speech API, Microsoft Speech Server, Amazon Alexa and Nuance Recognizer. These systems are marketed as commercial proprietary softwares for computers, smartphones or stand-alone devices for end-users; this offers very little control over the recognizer's features, and limited native integrability into other softwares, leading to release of open-source automatic speech recognition (ASR) systems. These are not meant for release as a market product but for developers; to understand, study and adapt according to the need of target application and use-case.

## 1.2  Statement of the Problems

Studying various models and working on open-source toolkits to convert speech to text. Then performing sentiment analysis and extracting keywords from this text. Thus, building ASR system for use-cases like customer reviews, telephonic queries, etc. Finally, evaluating and analysing the findings and results obtained from different components of the system.

## 1.3  Objectives of the Research

Everything we express (either verbally or in written) carries huge amounts of information. The topic we choose, our tone, our selection of words, everything adds some type of information that can be interpreted and value extracted from it. In theory, we can understand and even predict human behaviour using that information.

But there is a problem: one person may generate hundreds or thousands of words in a declaration, each sentence with its corresponding complexity. If you want to scale and analyse several hundreds, thousands or millions of people or declarations in a given geography, then the situation is unmanageable.

Data generated from conversations, declarations or even tweets are examples of unstructured data. Unstructured data doesn't fit neatly into the traditional row and column structure of relational databases, and represent the vast majority of data available in the actual world. It is messy and hard to manipulate. Nevertheless, thanks to the advances in disciplines like machine learning, a big revolution is going on regarding this topic[1].

Thus, we need a complete system to structure and analyse this data for extracting relevant information, to be used for various applications, reducing the manual work.

## 1.4  Scope

The present work focuses on working with available datasets. The dialects include only Indian-English and American-English. The work on text processing includes only unsupervised learning. These can be scaled, improved or altered according to different application requirement.

# 2 LITERATURE REVIEW

All modern descriptions of speech are to some degree probabilistic. That means that there are no certain boundaries between units, or between words. Speech to text translation and other applications of speech are never 100% correct. That idea is rather unusual as we usually work with deterministic systems, creating a lot of issues specific only to speech technology. Using artificial neural networks (ANNs), to improve speech-recognition performance, through a model known as the ANN-HMM have shown promise for large-vocabulary speech recognition systems achieving high recognition accuracy and low WER. Developing speech corpus depending upon the nature of language and addressing the issues of sources of variability through approaches like Missing Data Techniques & Convolutive Non-Negative Matrix Factorization[2], are the major considerations for developing an efficient ASR.

## 2.1 Approaches So Far

### 2.1.1 Acoustic-Phonetic

The method has been studied and used for more than 40 years, based upon theory of acoustic phonetics. This is based on identification of speech sounds and labelling these appropriately. It postulates that there exist finite, distinctive phonetic units in spoken language and these units are characterized by set of acoustic properties that keep changing in speech waveform over time. Hendal and Hughes[3] took the basis of finding speech sounds and providing labels to them and proposed that there exist a fixed number of distinctive phonetic units in spoken language which are broadly characterized by a set of acoustics properties varying with respect to time in a speech signal. For commercial applications, this approach has not provided a viable platform. This approach is implemented sequentially - spectral analysis, features detection, segmentation & labelling, and recognising valid words.

### 2.1.2 Pattern Recogniton

Itakura (1975)[4] was the first to propose this approach which got a considerable support from Rabiner and Juang (1989, 1993) to further the approach amongst researchers and developers. Pattern training and pattern recognition are two essential components of this method, where after a direction comparison is made between the word to be recognized from the speech, and possible pattern learned in the training stage for determination.

### 2.1.3 Knowledge Based

This requires expert knowledge about variations in speech, and is hand coded into the system. This approach gives the advantage of explicit modelling but this situation is difficult to obtain successfully. Knowledge based approach uses the information regarding linguistic, phonetic and spectrogram. The test speech is considered by all codebooks, and ASR chooses the word whose codebook yields the lowest distance measure.

### 2.1.4 SVM Classifier

Sendra et al.[5] have worked on a pure SVM-based continuous speech recogniser by applying SVM for making decisions at frame level and a Token Passing algorithm to obtain the chain of recognized words. The Token Passing Model is an extension of the Viterbi algorithm meant for continuous speech recognition so as to manage the uncertainty about the number of words in a sentence. The results achieved from the experiments have concluded that with a small database, recognition accuracy improves with SVMs but with a large database, the same result is obtained at the expense of huge computational effort.

## 3 METHODOLOGY

## 3.1 Concepts

### 3.1.1 Introduction to Speech

Speech is a complex phenomenon. People rarely understand how it is produced and perceived. The naive perception is often that speech is built with words and each word consists of phones; actually speech is dynamic and does not contain clearly distinguishable parts.

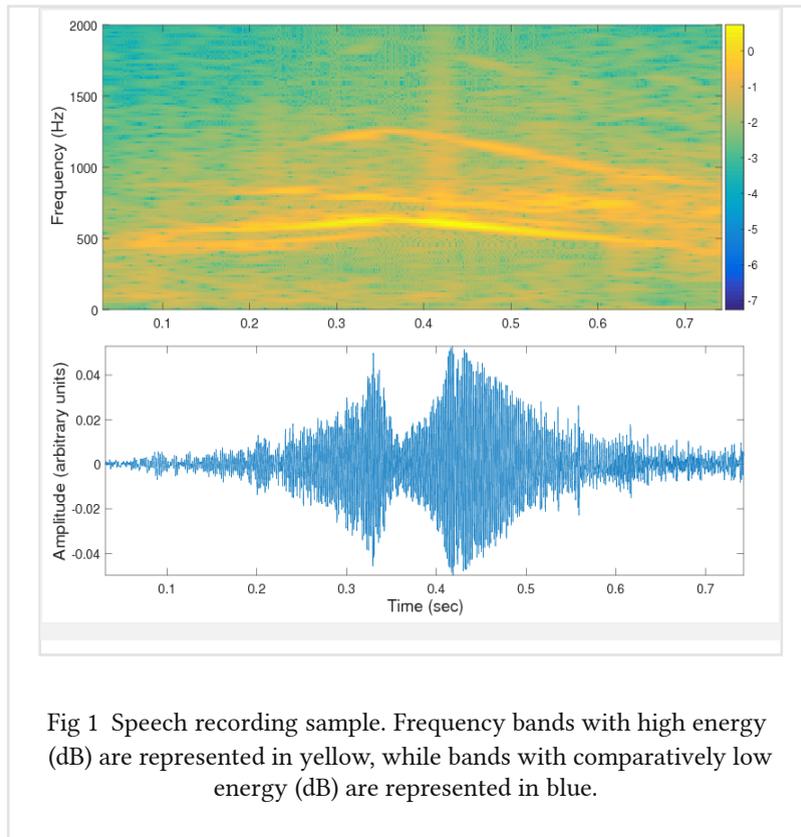

Fig 1 Speech recording sample. Frequency bands with high energy (dB) are represented in yellow, while bands with comparatively low energy (dB) are represented in blue.

It is a continuous audio stream where rather stable states mix with dynamically changed states. Words are understood to be built of phones, but this is certainly not true. The acoustic properties of a waveform corresponding to a phone can vary greatly depending on many factors - phone context, speaker, style of speech, etc. Transitions between words are often more informative than stable regions, therefore we often talk about diphones - parts between two consecutive phones.

The first part of the phone depends on its preceding phone, the middle part is stable and the next part depends on the subsequent phone - the three states in a phone selected for speech recognition.

Sometimes phones are considered in context. Such phones in context are called triphones. For example "a" with left phone "b" and right phone "d" in the word "bad" sounds a bit different than the same phone "a" with left phone "c" and right phone "n" in word "can". Please note that unlike diphones, they are matched with the same range in waveform as just phones. They just differ by name because they describe slightly different sounds.

For computational purpose it is helpful to detect parts of triphones instead of triphones as a whole, for example if one wants to create a detector for the beginning of a triphone and share it across many triphones. The whole variety of sound detectors can be represented by a small amount of distinct short sound detectors. Usually we use 4000 distinct short sound detectors to compose detectors for triphones. We call those detectors senones. A senone's dependence on context can be more complex than just the left and right context. It can be a rather complex function defined by a decision tree[6].

Next, phones build subword units, like syllables. For instance, when speech becomes fast, phones often change, but syllables remain the same.

Subwords form words, which are important in speech recognition because they restrict combinations of phones significantly. For instance, if there are 20 phones and an average word has 10 phones, there must be $20^{10}$ words; in practice, a person uses not more than 20 thousand[6].

Words and other non-linguistic sounds, which are called fillers (breath, um, uh, cough), form utterances. They are separate chunks of audio between pauses.

## 3.1.2  Recognition of Speech

We take a speech recording (waveform), split it at utterances by silences and then try to recognize what's being said in each utterance. To do that, we want to take all possible combinations of words and try to match them with the audio. We choose the best matching combination.

There are some important concepts in this matching process. First of all, it is the concept of features. Since the number of parameters is large, it needs optimization, by dividing the speech into frames. Then for each frame, typically of 10 milliseconds length, 39 numbers (generally) are extracted that represent the speech, which is called a feature vector.

**Feature Extraction**

In the ASR system features are extracted i.e. components which are useful and sufficient for identifying the linguistic content are used, and other redundant information is discarded, facilitating subsequent processes using the reduced representation of waveform.

- MFCC

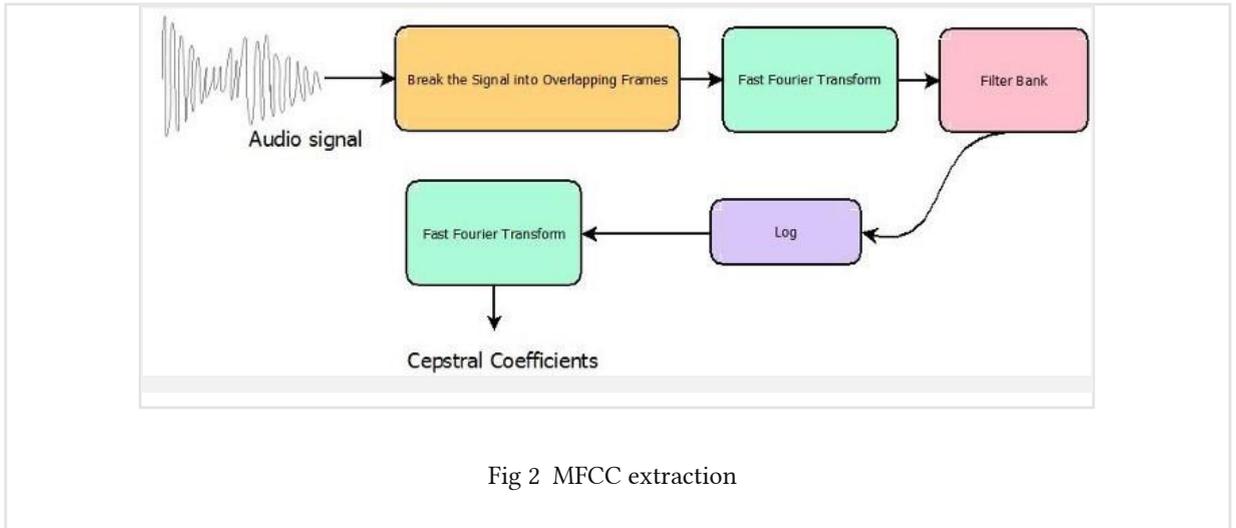

Fig 2  MFCC extraction

First we break the speech waveform into frames/samples. These are converted to frequency domain using FFT. On taking the log of the magnitude of the Fourier spectrum, and then again taking the spectrum of this log by a cosine, we observe a peak wherever there is a periodic element in the original time signal. Since we apply a transform on the frequency spectrum itself, the resulting spectrum is neither in the frequency domain nor in the time domain and hence called quefrency domain[7]. This spectrum of log of spectrum of time signal was named cepstrum. A frequency measured in Hertz (f) can be converted to the Mel scale using the following formula:

$$Mel(f) = 2595 \log(1 + \frac{f}{700})$$

$$( 1 )$$

Thus, MFCC features represent phonemes from speech accurately, for further processing, with the help of coefficients in the generated cepstrum.

• I-Vector

I-Vector is a vector of dimension several hundred (one or two hundred, in this particular context) which represents the speaker properties. Our idea is that the I-Vector gives the neural net as much as it needs to know about the speaker properties. This has proved quite useful. It is estimated in a left-to-right way, meaning that at a certain time t, it sees input from time zero to t. It also sees information from previous utterances of the current speaker, if available. The estimation is Maximum Likelihood, involving GMMs.

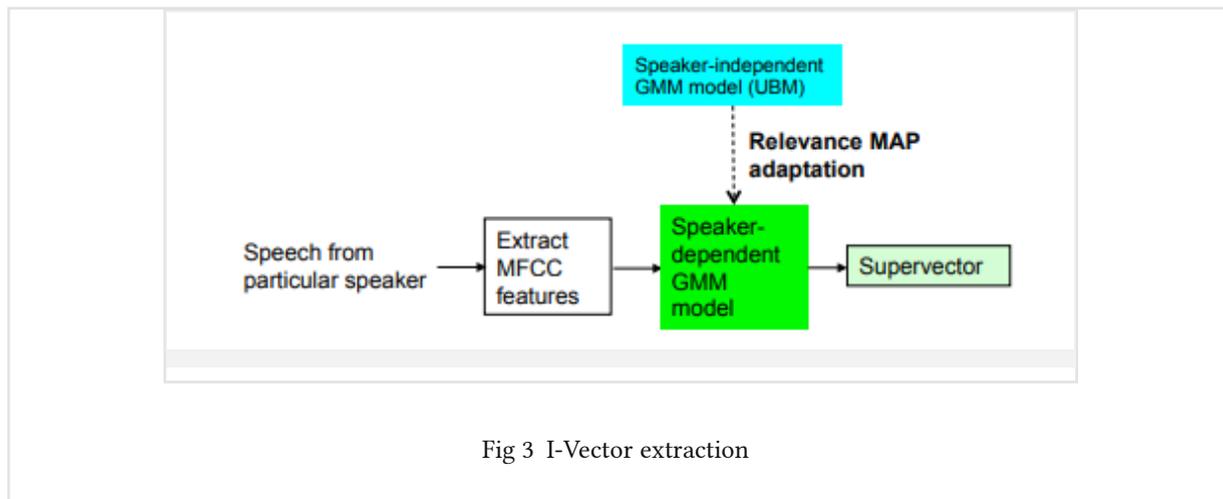

Fig 3  I-Vector extraction

A supervector for a speaker should be decomposable into speaker independent, speaker dependent, channel dependent, and residual components. Each component can be represented by a low-dimensional set of factors, which operate along the principal dimensions (i.e. eigendimensions) of the corresponding component. For instance, the following illustrates the speaker dependent component (known as the eigenvoice component) and corresponding factors:

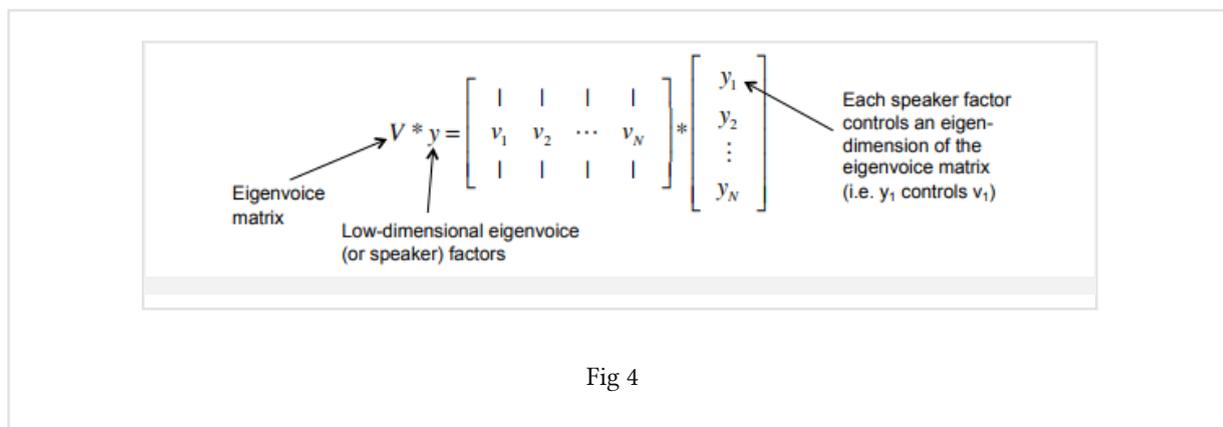

Fig 4

An I-Vector system uses a set of low-dimensional total variability factors (w) to represent each conversation side. Each factor controls an eigen-dimension of the total variability matrix (T), and are known as the I-Vectors[8].

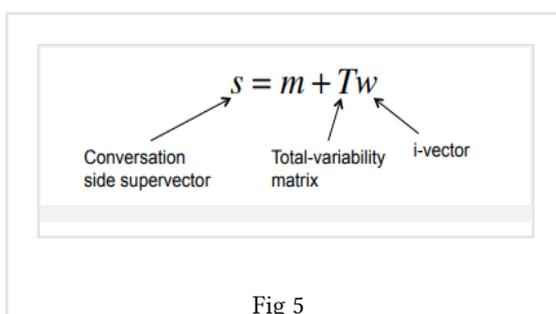

Fig 5

### 3.1.3 Acoustic Model

This contains the acoustic information about the speech for each senone. It specifies the relationship between the phone state at time t, $q_t$, and the acoustic spectrum, $x_t$. Since $x_t$ is a real-valued vector, instead of a symbol, the probability $P(x_t|q_t)$ must be a real-valued function. Most often, we model it using a weighted sum of Gaussians:

$$P(x_t|q) = \sum_{k=1}^{K} c_{qk} \prod_{d=1}^{D} \frac{1}{\sqrt{2\pi\sigma^2_{\text{qkd}}}} \; e^{-\frac{(x_{td} - \mu_{qkd})^2}{2\sigma^2_{\text{qkd}}}}$$

( 2 )

where, $x_{td}$ is the d$^{\text{th}}$ component of the vector $x_t$, and the parameters $c_{qk}$, $\mu_{qkd}$, and $\sigma^2{}_{qkd}$ have been learned from training data[9].

### 3.1.4 Language and Transition Model

This specifies the transition probability that the speech contains word "w4" given that it has previous words as (say) "w1", "w2" and "w3"; used to restrict word search and matching process. The most common language models are n-gram language models; these contain statistics of word sequences and finite state language models which define speech sequences by weights.

$$P(w_4|w_1, w_2, w_3) = \frac{N(w_1, w_2, w_3, w_4)}{N(w_1, w_2, w_3)}$$

( 3 )

where, N(w) is the frequency count of event w in a training database[9].

### 3.1.5 Phonetic Dictionary

The mapping of each phone from training data to a word is stored in the form of a phonetic dictionary. Each of these has different variants of pronunciation. Essentially all dictionaries have the same format: each line contains one word, followed by its pronunciation(s).

### 3.1.6 Decoder and Decoding Graph

The decoder accepts the above three, along with the speech waveform, as input and transcribes to give the output or hypothesis.

A directed graph represents variants of the recognition called a lattice. Often, getting the best match is not practical or required. In that case, lattices are good intermediate formats to represent the recognition result.

**Veterbi Algorithm**

This is used for finding the most probable sequence of hidden states (veterbi path) that results in a sequence of observed events; which is sequence of words in speech recognition in polynomial time.

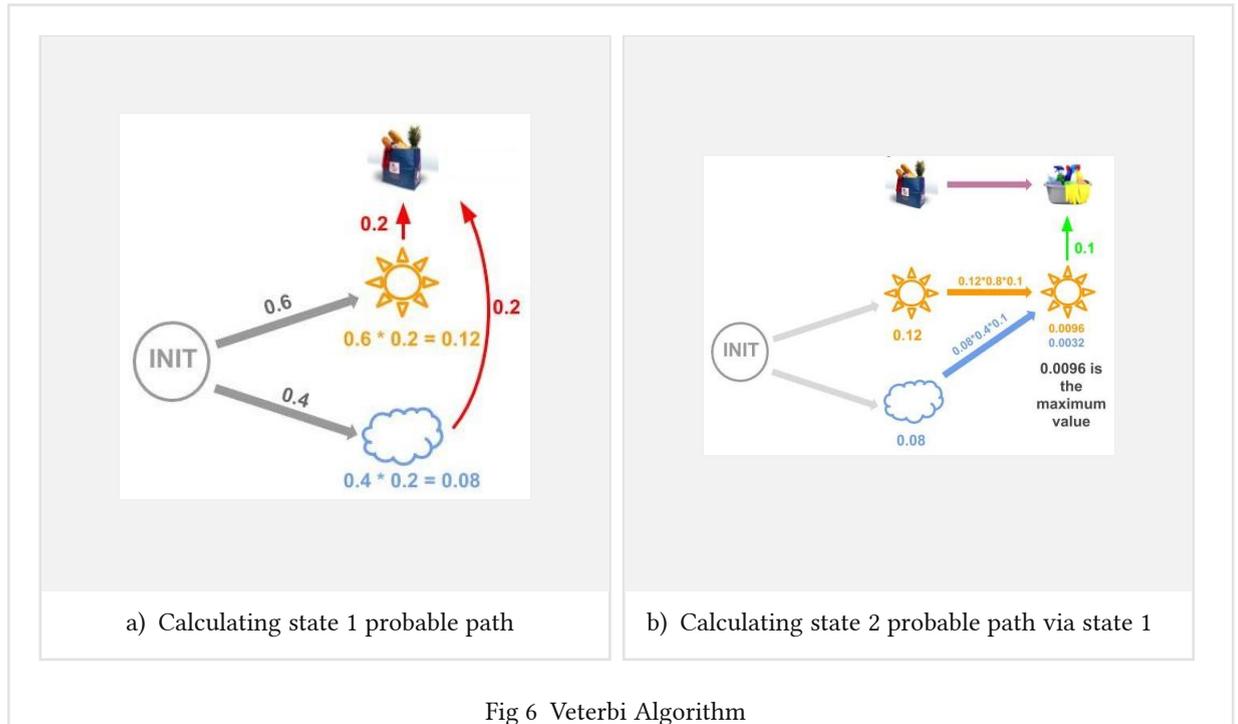

|  |  |
|---|---|
| a) Calculating state 1 probable path | b) Calculating state 2 probable path via state 1 |

Fig 6  Veterbi Algorithm

## 3.2 Methods

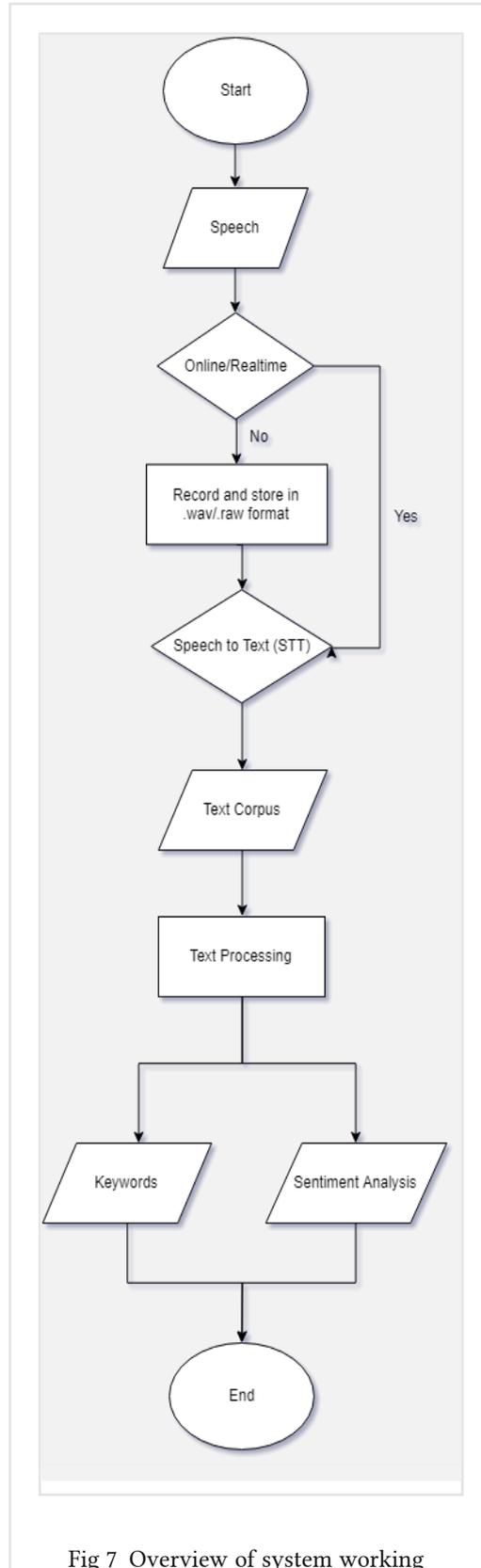

Fig 7  Overview of system working

## 3.2.1 Experimental Setups

- System Configuration: Intel i5 7$^{th}$ Gen., 8 GB RAM

- Operating Systems: Linux Ubuntu 16.04, Windows 10

- Programming Language: Python 2.7, Python 3.7

- Toolkits: SphinxBase, SphinxTrain, PocketSphinx, pyKaldi ASR

- APIs and Libraries: pyAudio, pke, sounddevice, soundfile, NLTK, Spacy, TextBlob

- Other Softwares: Bash, Git, The ocrevalUAtion tool

## 3.2.2 Speech to Text

The toolkits are used with HMM - GMM (CMUSphinx) and HMM - DNN (Kaldi) acoustic models, ARPA language models and English phonetic dictionary (lexicon).

If X denotes sequence of acoustic feature vectors and W denotes word sequence, the most probable word sequence W* will be given by:

$$\mathbf{W}^* = \arg\max_{\mathbf{W}} P(\mathbf{W}\,|\,\mathbf{X})$$

Applying Bayes' Theorem:

$$P(\mathbf{W}\,|\,\mathbf{X}) = \frac{p(\mathbf{X}\,|\,\mathbf{W})\,P(\mathbf{W})}{p(\mathbf{X})}$$

$$\propto p(\mathbf{X}\,|\,\mathbf{W})\,P(\mathbf{W})$$

$$\mathbf{W}^* = \arg\max_{\mathbf{W}} \underbrace{p(\mathbf{X}\,|\,\mathbf{W})}_{\text{Acoustic model}} \underbrace{P(\mathbf{W})}_{\text{Language model}}$$

Fig 8  HMM

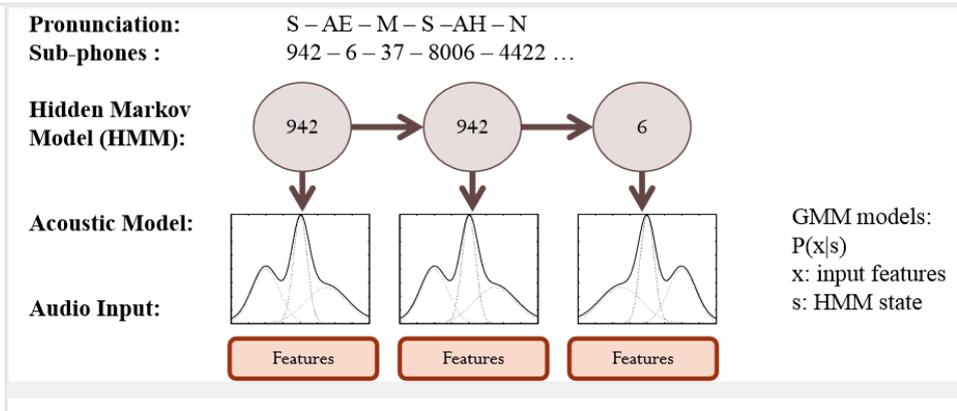

Fig 9  HMM - GMM

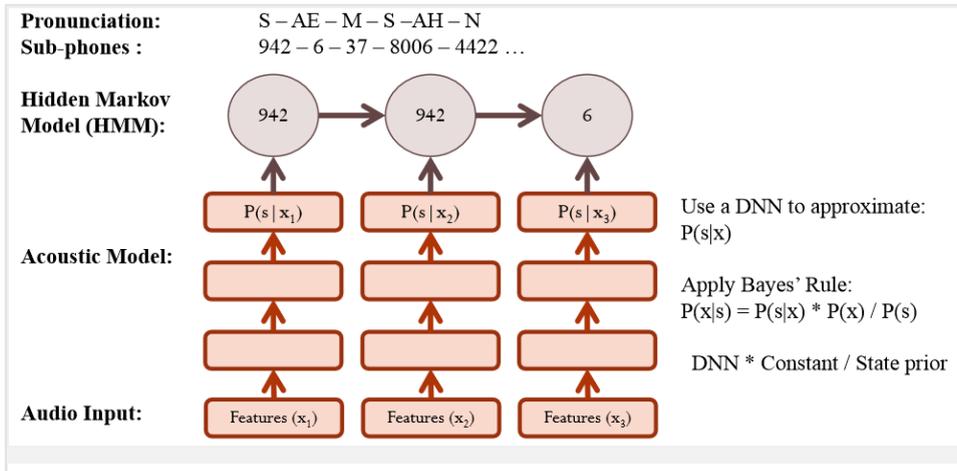

Fig 10  HMM - DNN

```
 1. \data\
 2. ngram 1=7
 3. ngram 2=7
 4.
 5. \1-grams:
 6. -1.0000 <unk>   -0.2553
 7. -98.9366 <s>     -0.3064
 8. -1.0000 </s>      0.0000
 9. -0.6990 hello -0.2553
10. -0.6990 this -0.2553
11. -0.6990 is -0.2553
12. -0.6990 example -0.1973
```

```
13.
14. \2-grams:
15. -0.2553 <unk> wood
16. -0.2553 <s> <unk>
17. -0.2553 hello this
18. -0.2553 is example </s>
19.
20. \end\
```

Code 1  Sample trained ARPA language model

```
1. hello H EH L OW
2. world W ER L D
```

Code 2  Sample phonetic dictionary excerpt

The models are converted to Weighted Finite State Transducer (WFST) during the decoding recipe "HCLG.fst" in Kaldi.

H.fst: This maps multiple HMM states (a.k.a. transition-ids) to context-dependent triphones i.e. on the right are the context-dependent phones, and on the left are the pdf-ids.

C.fst: This maps triphone sequences to monophones (senones).

L.fst: The file is the Finite State Transducer form of the lexicon with phones on the input side and words on the output side.

G.fst: This is the grammar (can be built from an n-gram ARPA language model).

## 3.2.3  Keywords Extraction

NLTK - RAKE: A list of candidate keywords/phrases (without stopwords) is made. A Co-occurrence graph is built to identify the frequency of words associated together in those phrases[10]. An individual word score is calculated as the degree (number of times it appears + number of additional words it appears with) of a word divided by it's frequency (number of times it appears), which weights towards longer phrases. Adjoining stopword is included if they occur more than twice in the document and score high enough.

TextRank: It creates a graph of the words and relationships between them from the text, then identifies the most important vertices of the graph (i.e. words) based on importance scores calculated recursively from the entire graph.

TopicRank: The document is preprocessed (sentence segmentation, word tokenization and POS tagging) and keyphrase candidates are clustered into topics. Then, topics are ranked according to their importance in the document, and keyphrases are extracted by selecting one keyphrase candidate for each of the most important topics[11].

YAKE: First is the pre-processing step which splits the text into individual terms whenever an empty space or a special character, like line breaks, brackets, comma, period, etc., delimiter is found. Second, we devise a set of five features to capture the characteristics of each individual term. These are - (1) Casing; (2) Word Positional; (3) Word Frequency; (4) Word Relatedness to Context; and (5) Word DifSentence.

TF - IDF: Finds the words that have the highest ratio of occurring in the current document vs the frequency of occurring in the larger set of documents.

### 3.2.4 Sentiment Analysis

TextBlob is a Python library for processing text. It provides an API for NLP tasks such as part-of-speech tagging, noun phrase extraction, sentiment analysis, classification, translation, etc.

# 4 RESULTS AND DISCUSSIONS

## 4.1 Speech to Text

### 4.1.1 CMUSphinx

- Dataset

Recorded files of the format .raw (16KHz) with approximately 10,000 spoken characters in the form of independent sentences (American-English - Model 1 and Model 2, Indian-English - Model 3) were used, along with the transcription files of the format .txt to test different models.

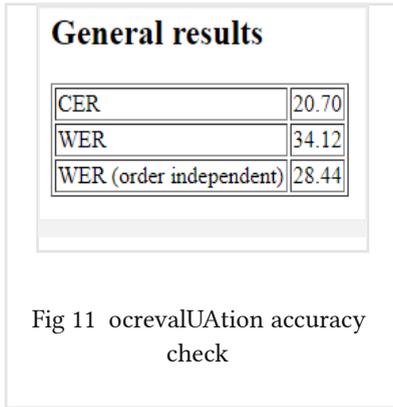

Fig 11  ocrevalUAtion accuracy check

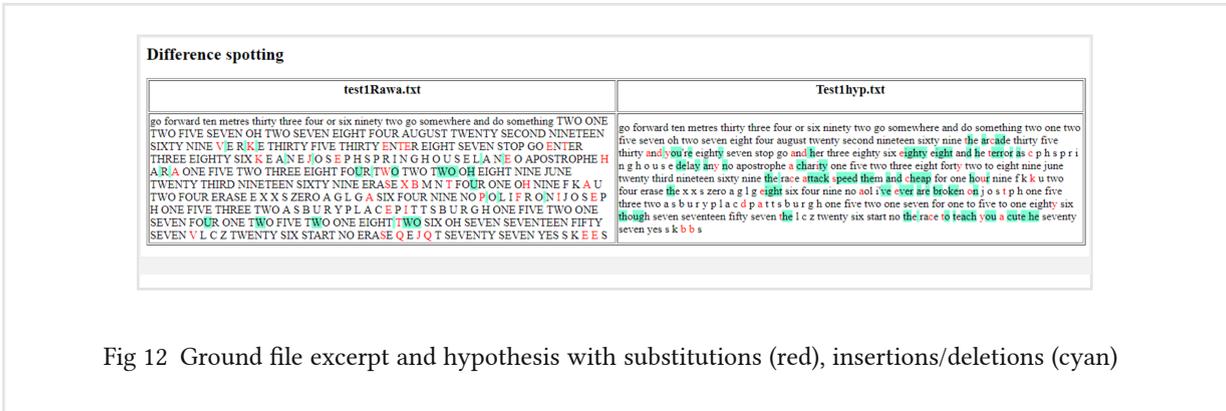

Fig 12  Ground file excerpt and hypothesis with substitutions (red), insertions/deletions (cyan)

- Model 2

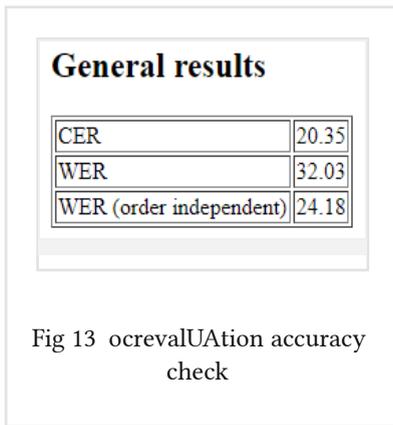

Fig 13  ocrevalUAtion accuracy check

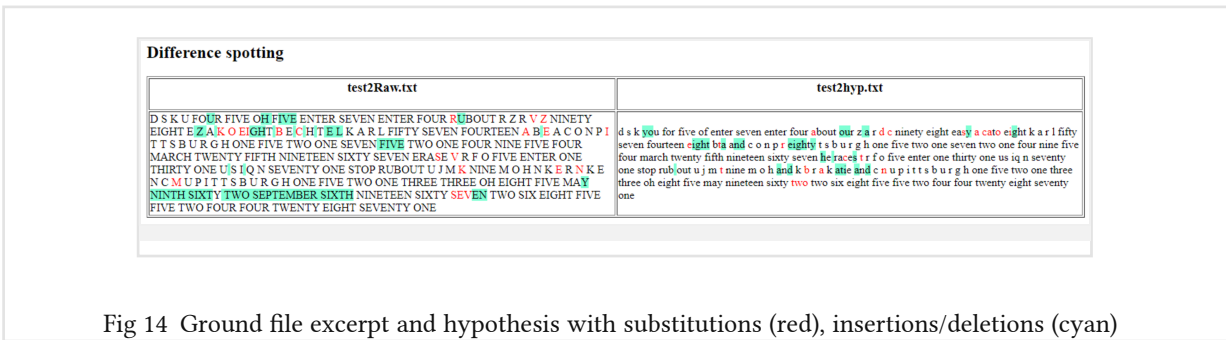

Fig 14  Ground file excerpt and hypothesis with substitutions (red), insertions/deletions (cyan)

- Model 3

**General results**

| CER | 66.67 |
|---|---|
| WER | 78.26 |
| WER (order independent) | 73.91 |

Fig 15  ocrevalUAtion accuracy check

Fig 16  Ground file excerpt and hypothesis with substitutions (red), insertions/deletions (cyan)

The models perform better with Armerican-English accent as compared to Indian-English, with Model 2 outperforming the other two.

## 4.1.2  Kaldi

- Dataset

Recorded files of the format .wav (16KHz) with approximately 10,000 spoken characters in the form of independent sentences (Indian-English) were used, along with the transcription files of the format .txt to test different models.

- Aspire Chain Model

**General results**

| CER | 51.69 |
|---|---|
| WER | 66.95 |
| WER (order independent) | 54.60 |

Fig 17  ocrevalUAtion accuracy check

**Difference spotting**

| hyp.txt | decode.txt |
|---|---|

Fig 18  Ground file excerpt and hypothesis with substitutions (red), insertions/deletions (cyan)

- Zamia Speech Model

**General results**

| CER | 12.34 |
|---|---|
| WER | 21.64 |
| WER (order independent) | 18.42 |

Fig 19  ocrevalUAtion accuracy check

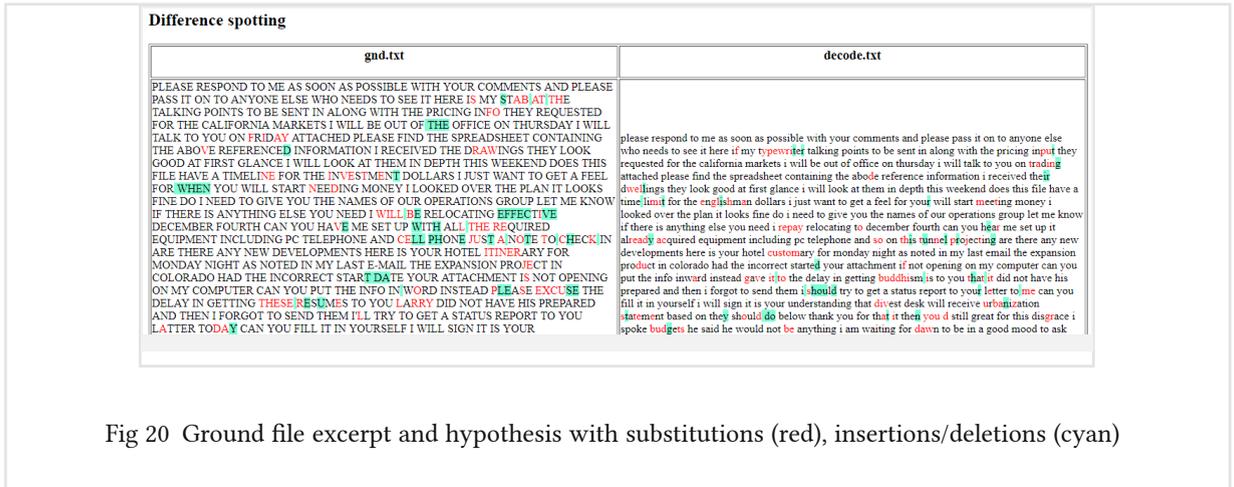

Fig 20  Ground file excerpt and hypothesis with substitutions (red), insertions/deletions (cyan)

The models give better results on Indian-English accent as compared to those used with CMUSphinx, with Zamia Speech Model giving the best results.

# 4.2  Keywords Extraction

- Dataset 1

Movie reviews and gists from IMDb and Rotten Tomatoes with approximately 10,00,000 characters segregated into two files as positive and negative.

Table 1  Comparison of time taken to analyse the input

| NLTK-RAKE | N/A |
| --- | --- |
| TextRank | 97.96 seconds |
| TopicRank | N/A |
| YAKE | 153.69 seconds |
| TF-IDF | N/A |

Only TextRank and YAKE could give relevant and appropriate results, while other algorithms either gave arbitrary phrases or could not process the text.

TextRank gave phrases like "feature film", "family film", "very good natured", etc. (Positive.txt) and "old", "bad film", etc. (Negative.txt).

YAKE gave phrases like "one of the best", "very best", "good scenes", etc. (Positive.txt) and "not for the audience", "did not like the movie", etc. (Negative.txt).

- Dataset 2

Article on general information about India containig approximately 5,000 characters.

Table 2 Comparison of time taken to analyse the input

| | |
|---|---|
| NLTK-RAKE | 0.01 seconds |
| TextRank | 0.55 seconds |
| TopicRank | 0.59 seconds |
| YAKE | 0.61 seconds |
| TF-IDF | 2.42 seconds |

For the smaller corpus, all the algorithms performed well; RAKE, being the fastest to process (0.01 seconds) gave most relevant results with phrases like "famous tourist country", "populated", "unity in diversity", "taj mahal", "qutub minar", "religions", etc.

## 4.3 Sentiment Analysis

- Dataset 1

Movie reviews and gists from IMDb and Rotten Tomatoes with approximately 10,00,000 characters segregated into two files as positive and negative.

Table 3 Positive.txt

| | |
|---|---|
| Polarity | 13.67% (positive) |
| Subjectivity | 51.42% |
| Time | 1.36 seconds |

Table 4 Negative.txt

| | |
|---|---|
| Polarity | 4.95% (negative) |
| Subjectivity | 50.93% |

| Time | 1.27 seconds |
|------|--------------|

- Dataset 2

Article on general information about India containig approximately 5,000 characters.

Table 5 India.txt

| Polarity | 28.98% (positive) |
|----------|-------------------|
| Subjectivity | 54.78% |
| Time | 0.07 seconds |

# 5  CONCLUSION AND FUTURE WORK

The system to recognize speech and process it for extracting keywords and sentiment analysis was worked upon. Various combinations of toolkits, models, libraries and algorithms were used to analyse the different components like STT, text processing, etc., for accuracy, speed and use-cases. Further, models can be trained with datasets according to the need and application, and tested with various approaches mentioned to get desired results for developing a complete NLP system.